\documentclass{ws-ijgmmp}

\begin{document}


%
\catchline{}{}{}{}{}
%
\def\beq{\begin{equation}}
\def\be{\begin{equation}}
\def\eeq{\end{equation}}
\def\ee{\end{equation}}
\def\bea{\begin{eqnarray}}
\def\eea{\end{eqnarray}}

\title{Effective field theory from modified gravity with massive modes
}

\author{Salvatore Capozziello}
\address{Dipartimento di Fisica, Universit\`{a} di Napoli ``Federico II'', Napoli, Italy.\\
Istituto Nazionale di Fisica Nucleare (INFN) Sez. di Napoli, Compl. Univ. di Monte S. Angelo, Edificio G, Via Cinthia, I-80126, Napoli, Italy.\\
Gran Sasso Science Institute (INFN), Via F. Crispi 7, I-67100, L' Aquila, Italy.\\
}

\author{Mariafelicia De Laurentis}
\address{Tomsk State Pedagogical University, Department of Theoretical Physics, \\ pr. Komsomolsky, 75, Tomsk, 634041, Russia \\
}

\author{Mariacristina Paolella}

\address{Dipartimento di Fisica, Universit\`{a} di Napoli ``Federico II'', Napoli, Italy.\\
Istituto Nazionale di Fisica Nucleare (INFN) Sez. di Napoli, Compl. Univ. di Monte S. Angelo, Edificio G, Via Cinthia, I-80126, Napoli, Italy.\
 }

\author{Giulia Ricciardi}
\address{Dipartimento di Fisica, Universit\`{a} di Napoli ``Federico II'', Napoli, Italy.\\
Istituto Nazionale di Fisica Nucleare (INFN) Sez. di Napoli, Compl. Univ. di Monte S. Angelo, Edificio N, Via Cinthia, I-80126, Napoli, Italy.\
 }

\maketitle

\begin{history}
\end{history}

\begin{abstract}
Massive gravitational modes  in effective field theories can be recovered by extending General Relativity and  taking into account  generic functions of the curvature invariants,  not necessarily linear in the Ricci scalar $R$. In particular, adopting the minimal extension of $f(R)$ gravity, an effective field theory with  massive modes is straightforwardly recovered. This approach allows to  evade shortcomings like ghosts and discontinuities if a suitable choice of expansion parameters is performed. 
\end{abstract}


\section{Introduction}

The long standing problem of  graviton mass  \cite{Fierz:1939ix,vanDam:1970vg,Zakharov,Vainshtein1972,Boulware:1973my,Arkani-Hamed:2002sp,Rubakov:2008nh} has recently excited  a renewed  interest both at fundamental  and  cosmological level.  From one side, massive gravitational states  could be the signature  of  some effective theory quantization. From the other side, massive gravitons could be  the natural candidates  for dark matter   capable of structuring self-gravitating astrophysical  systems \cite{wonlee,casadio}.

Even though a  quantum  description of gravity has not  been
 achieved yet \cite{Smolin03, Kiefer06},   it is possible to quantize gravity in the  linear approximation of Minkowskian limit. Specifically,
assuming General Relativity (GR) as the theory of gravitational interaction,   the quantization in this limit    gives rise to    spin-2 massless bosons, i.e. the massless gravitons.  Starting from this result,   a reasonable question to ask is  whether  gravitons could be  massive in some alternative theory of gravity where GR is only a limiting or a particular case  \cite{Rubakov:2008nh, hinterbichler}.
 However, the concept of  massive gravitons poses some controversial issues   that greatly complicate the formulation of self-consistent  theories, such as the presence of ghost, instabilities, discontinuity and strong coupling effects at low energy scales \cite{vanDam:1970vg,Zakharov,Boulware:1973my,Arkani-Hamed:2002sp}.
In any case,   massive graviton solutions cannot be simply ruled
out if one wants to face coherently the problem of gravitational interaction in the ultraviolet limit \cite{Vainshtein1972, Visser98, DDGV02, FS02}.

On the other hand,   a large amount of  alternative theories of  gravity has been recently developed in order to match the problem of  cosmic acceleration in view of dark energy  \cite{PhysRepnostro,OdintsovPR,Mauro,DeFelice, DGP00}.   In all these  approaches,  the problem of  massive gravitons emerges and has to be consistently considered also at infrared limit \cite{greci,basini}.
The original motivation is related to the observational evidence of the accelerated expansion of the Universe at  the
 present epoch.
This accelerated expansion may be due to the cosmological constant, to new weakly interacting fields constituting  some kind of dark energy. The problem is that the  dark energy scale appears to be smaller and smaller with respect to   the energy scale of any  known interactions. The unnatural smallness of  dark energy density constitutes  the cosmological constant problem.
In this sense,
infrared-modified gravity models could be phenomenologically relevant as a possible alternative to dark matter and
dark energy whose effects at large scales  could be  originated by geometry, specifically by the further  degrees of freedom emerging in alternative theories of gravity \cite{Mauro}.

Based on the previous, as well as other, motivations,
there have been several experimental searches for massive gravitons,
resulting in  upper limits for the mass which differ by several orders of magnitude.  For example, a  limit on  the graviton mass   ($\sim 8\times 10^{4}$ eV) has been achieved  by measuring the decay of two photons  \cite{Hare73}. Besides, assuming that clusters of galaxies are bound by more-or-less standard
gravity, it is possible to obtain an upper limit of $2\times 10^{-29} h_{0}$ eV, where
$h_0$ is the Hubble constant in units of 100\,km\,s$^{-1}$\,Mpc$^{-1}$ \cite{GN74}.

On the other hand, gravitational waves sector has a prominent role in this discussion. Gravitational waves coming from GR
are described by the transverse-traceless gauge, which is a spin-$2$ tensor under rotations with massless modes.  Beside of these standard results, it is possible to construct consistent models where Lorentz invariance is broken and the masses of scalar, vector and tensor perturbations are different from zero. A direct limit on the mass of graviton  can be obtained from gravitational waves
by  binary stellar systems and from the  inspiral rate   inferred from the timing
of binary pulsars. This bound is about  $7.6 \times 10^{-20}$ eV for the  binary pulsar PSR B1913+16 \cite{FS02}. The same limit can be also obtained by studying  binary systems in $f(R)$-gravity \cite{ivan}. An estimate of the  graviton mass upper limit  of about $ 7 \times 10 ^ {-32} $
eV, is obtained by considering the effect of  gravitons on
the power spectrum of weak lensing, with assumptions about dark energy
and other parameters  \cite{Will06}.

From a  genuine theoretical point of view, the study of gravitons is challenging  due to the
 problem of reconciling gravity and   quantum field theories describing fundamental interactions.
A bridge  is represented  by effective field theories (EFT), that allow  to analyze  different energy regimes separately (see e.g. \cite{Donoghue:1995cz, Burgess:2003jk}).
In general,  since the effective Lagrangians is
 non-renormalizable, due to  an infinite number of counterterms, one retains  only a few of them,  in a phenomenological approach where only leading terms are necessary. This means that the
 determination of the effective degrees of freedom is a crucial point for any effective theory and this fact  is even more important in connection with gravity.

Technically, a way to build up  an effective Lagrangian
is to identify some expansion  parameters  and
classify terms in the Lagrangian according to such  parameters. Without knowing  the underlying fundamental theory, the coefficients of the expansion are necessarily unknown, and their values have to be determined, in principle,  by experiments.

In this paper, we take into account an effective theory of  gravity that follows naturally from  Extended Theories of gravity (ETG) (see e.g. \cite{PhysRepnostro}).
The  action can be  expanded in powers of the Ricci curvature scalar $R$ satisfying a massive Klein Gordon field equation. In particular,
by linearizing  $f(R)$ gravity, the Lagrangian describes  a massive scalar field  where  a  mass scale  $m$  emerges naturally. The theory does not predict the value of this mass, but it does predict its connection with parameters of the ETG Lagrangian. It is possible to identify  correlations between the coefficients of the effective Lagrangioan, which  may,  in turn, induce correlations
among observables at different scales.
A first  result is that the assumption  of  an effective Lagrangian derived from $f(R)$ gravity  allows to escape the problem of scalar ghosts in massive theories,  as pointed out in  \cite{vanDam:1970vg}. In the limit where $m \gg \Lambda$ (being $\Lambda$  the cosmological  constant),  we achieve a physically acceptable  scalar field satisfying a homogeneous Klein Gordon equation and then one achieves an effective field Lagrangian bypassing some of the problems raised in  \cite{vanDam:1970vg} where  GR, i.e. $f(R)=R$, was considered.
The paper is organized as  follows. Sec. 2 is devoted to the construction of the field equations for  analytical  $f(R)$ gravity models. Here we put in evidence the main parameters of the theory defining an effective massive mode related to the further degrees of freedom coming from $f(R)$. The linearized theory is discussed in Sec.3. In particular,  we derive  and discuss the emergence of massive modes and how they contribute in the construction of the effective Lagrangian. The effective Lagrangian and its features are considered  in Sec. 4. Conclusions are drawn in Sec. 5.

\section{Field equations for  $f(R)$ gravity}
\label{sect3}

Let us   consider  a $4$-dimensional action in vacuum for a generic function $f(R)$ of the Ricci scalar \cite{PhysRepnostro,OdintsovPR,Mauro,DeFelice}
\begin{equation}
S= \int d^4x \sqrt{-g} f(R)\,,
\label{LagrangianF}
\end{equation}
 where the Ricci scalar is defined as $R=g^{\mu\nu}R_{\mu\nu}$, and $g$ is the determinant of the metric.
 The only  assumption at this stage  is  that $f(R)$ is an analytic function (\textit{i.e.} Taylor expandable) in term of the
Ricci scalar,  that is
 \begin{equation}
f(R)=\sum_{n}\frac{f^{n}(R_0)}{n!} (R-R_0)^n= f_0 + f_0^{'} R+ \frac{1}{2} f''_0 R^2+...\,,
\end{equation}
where we recover the flat-Minkowski background as soon as  $R=R_0=0$ and $f_0=0$.
 Here  ${\displaystyle f'(R)=\frac{df(R)}{dR}}$ and  ${\displaystyle f''(R)=\frac{d^2f(R)}{dR^2}}$ indicate the derivative with respect to
 the Ricci scalar $R$. We have defined $f_0 = f(R)|_{R=R_0}$, $ f_0^{'} =  f'(R)|_{R=R_0}$    and so on.
At the second order of approximation in term of $R$, the above action (\ref{LagrangianF}) becomes:
 \begin{equation}
 S=
  \int d^4x \sqrt{-g} \left[f_0 + f'_0 R+\frac{1}{2} f''_0 R^2\right]\,.
\label{S1}
 \end{equation}
This can be viewed as an EFT Lagrangian,  naturally emerging in the context of ETG. In a bottom-up approach, from the point of view of unconstrained EFT,
 there is no rationale, like symmetries or renormalizability, for choosing the gravitational action proportional to $R$ like in GR, except indications that the curvature $R$ is rather small. Moreover, there are infinite  terms allowed by general coordinate invariance, such as $R_{\mu\nu} R^{\mu\nu}$, where $ R_{\mu\nu} $ is the  Ricci tensor, $R_{\mu\nu\lambda\sigma} R^{\mu\nu\lambda\sigma} $, where $R_{\mu\nu\lambda\sigma}$ is the  Riemann tensor, derivatives of $R$, and so on. Where one has to  truncate the expansion is somehow a matter of choice, and the coefficients are completely unknown from a theoretical point of view. Instead, the terms in the action \eqref{S1}
follow from the underlying ETG, which can also give indications on  the coefficients and the order of the series.  Here we are choosing the simplest possibility considering an analytical $f(R)$ theory of gravity.

By varying  the action \eqref{S1} with respect to the metric, we obtain the field equations
\begin{equation}
-\frac{f_0}{2} g_{\mu \nu}+ f'_0 G_{\mu \nu} -
f''_0 \left[  \nabla_\mu \nabla_\nu R - g_{\mu \nu} \Box R
+ R
\left( \frac{1}{4} R g_{\mu \nu} - R_{\mu \nu} \right)    \right] = 0\,,
\label{FE}
\end{equation}
where
\begin{equation}
G_{\mu \nu}=R_{\mu\nu}-\frac{1}{2}g_{\mu\nu}R
\end{equation}
 is the Einstein tensor and
 $\displaystyle{\Box={\nabla_\sigma \nabla^\sigma}} $ is the d'Alembert operator with $\nabla_{\sigma}$  indicating  covariant derivatives.
It is interesting to note that if we
rewrite the ETG  Lagrangian in the form:
  \begin{equation}
\mathcal{L}= \sqrt{-g} \left[\frac{f_0}{f'_0} + R+\frac{1}{2} \frac{f_0^{''}}{f_0^{'}} R^2\right]\, f^\prime_0,
\label{lagrangian}
 \end{equation}
we can identify the cosmological constant term as  $\displaystyle{\frac{f_0}{f_0^{'}} =-2\Lambda }$. We are working in Planck units, therefore we  assume that the Lagrangian in Action  \eqref{lagrangian} is multiplied by $1/16 \pi G$, where $G$ is the Newton constant. From now on, we will work in  in "modified" Planck units, that is we will assume a multiplicative factor  $1/16 \pi \tilde G$,  with $\tilde G=G/f^\prime_0$, that reduces to the standard one as soon as
$f^\prime_0=1$.
Immediately, we have
\begin{equation}
\Lambda g_{\mu \nu}+ G_{\mu \nu} -\frac{f''_0}{f'_0}
 \left[
 \nabla_\mu \nabla_\nu R
 - g_{\mu \nu} \Box R
+ R \left( \frac{1}{4} R g_{\mu \nu} - R_{\mu \nu} \right)    \right] = 0\,.
\end{equation}
The trace of above field equations gives
\begin{equation}
\label{K-G+L}
\Box R - \frac{ f'_0} {3f''_0} (R - 4 \Lambda)= 0\,.
\end{equation}
Obviously, setting $f_0 =0$, that is discarding the $0^{th}$ term, is equivalent to set to zero the cosmological constant, and the trace
equation becomes
\begin{equation}
\label{K-G}
\Box R - \frac{ f'_0} {3f''_0} R= 0\,.
\end{equation}
 Eqs. (\ref{K-G+L}) and (\ref{K-G}) are  Klein-Gordon-like equations; indeed, by assuming that the ratio $ f_0^{'}/f_0^{''}$ is  negative,
we can define an effective  mass
\begin{equation}   m^2 \equiv -\frac{ f_0^{'}} {3f_0^{''}} \end{equation}
 so that we have
\begin{subequations}
\begin{align}
\label{eqK-Gfin}
&\Box R +m^2 R= 0\,, \\
\label{eqK-Gfin-b}
&\Box R +m^2 (R - 4 \Lambda)=0\,.
\end{align}
\end{subequations}
It follows  that the curvature $R$ can be considered formally analogous to a massive scalar field \cite{starobinsky}.
We can neglect the non-homogeneous equation as soon as  the condition
\beq
R \gg \Lambda
\eeq
holds. Let us study now the linearized version of such a theory in order  to interpret it in the context of  EFT.


\section{Linearized $f(R)$ gravity}

 In order to linearize  the field equations (\ref{FE}) at  first order in $h_{\mu\nu}$, we have to  expand  around the flat spacetime metric $\eta_{\mu \nu}$ \cite{Maggiore,DeLaurentis:2011re,DeLaurentis:2011tp}.
Therefore we have


\begin{equation}
g_{\mu \nu}=  \eta_{\mu \nu}+ h_{\mu \nu}\,, \qquad \Rightarrow \qquad ds^2= g_{\mu \nu} dx^{\mu}dx^{\nu}=
(\eta_{\mu \nu}+ h_{\mu \nu} )dx^{\mu}dx^{\nu}\,,
\end{equation}
with $h_{\mu \nu}$ small ($\mathcal{O} (h^2) \ll 1$). It is important to stress that the perturbation $h_{\mu\nu}$ is a symmetric tensor.
The  Ricci scalar, at the first order in  metric perturbation, reads
\begin{equation}
R=\partial^\sigma \partial^\tau h_{\sigma \tau} - \Box h\,,
\end{equation}
where $h \equiv  h^{\mu}_{\mu}$ is the trace of $h_{\mu\nu}$ and $\Box = \partial_{\sigma}\partial^{\sigma}$ that is reduced now to the standard d'Alembert operator defined on the underlying Minkowski spacetime where gravity is assumed as a perturbation.
Considering the  harmonic gauge condition\footnote{Such a condition is also called  Hilbert, or  De Donder  or Lorentz gauge. In  general, the harmonic  gauge is defined in a curved background  by the condition \
$ \partial_\nu \left(g^{\mu\nu}\sqrt{-g}\right)=0$. Writing  $g_{\mu \nu}=  \eta_{\mu \nu}+ h_{\mu \nu}$ and expanding to linear order, the harmonic gauge reduces to the standard Lorentz gauge.}  \beq {\partial^\mu h_{\mu \nu}}=0 \label{harm} \eeq  we find
\begin{equation}
R= - \Box h\,.
\label{Rh}
\end{equation}
The fluctuation of the metric on the background represents, in this approach,   the field mediating the gravitational  interaction.
 Our aim is now to identify its properties  by setting the corresponding field equations.

Let us consider  the homogeneous Klein-Gordon  Eq. (\ref{eqK-Gfin}).
Substituting the expression for $R$ given by Eq. \eqref{Rh}, we find
\begin{equation}
\label{boxh}
\Box (\Box h + m^2 h)= 0\,,
\end{equation}
%
 We can choose the trivial solution
\begin{equation}
\Box h + m^2  h= 0\,,
\end{equation}
and find the condition
\begin{equation}
\label{condition}
\Box h =- m^2 h\,,
\end{equation}
that is a sort of mass shell condition.
We can also consider Eq. \eqref{eqK-Gfin-b} discussing the role of cosmological constant. As it is well known, a general  solution is the sum of the field satisfying the associated homogeneous Eq.  (\ref{eqK-Gfin}) plus a particular solution  $R^\prime$, that we can formally write as
\beq R^\prime(x) =  4 \Lambda m^2 \int G(x, x^\prime) dx^\prime \eeq
Here $ G(x,x^\prime)$ is a non-local Green function satisfying the field equation
\begin{equation}
\label{boxh}
(\Box  + m^2) G(x, x^\prime)= \delta(x, x^\prime)\,.
\end{equation}
Only the  scale $m^2$ appears in  Eqs. \eqref{condition} and \eqref{boxh},  while $R^\prime$ is suppressed by $\Lambda$. We can reasonably assume that $R^\prime$ can be neglected with respect to the solutions of Eq.  (\ref{eqK-Gfin}), as far as the approximation $\Lambda \ll m^2$ holds.

Let us  now rewrite  the Lagrangian (\ref{lagrangian}) in term of the perturbation. It is
\begin{equation}
\mathcal{L} =\sqrt{- g} \; \left[\frac{f_0}{f'_0}+ R + \frac{f_0^{''}}{2 f'_0} R^2\right]= \sqrt{- \mathrm{det} \left(\eta_{\mu
\nu} + h_{\mu \nu} \right)} \left(R-2 \Lambda - \frac{1}{6 m^2} R^2\right)\,.
\label{EFT33}
\end{equation}
where  we have indicated the determinant of the metric.
Substituting $R \rightarrow -\Box h$, we find
\begin{equation}
\mathcal{L} = \sqrt{- \mathrm{det}  (\eta_{\mu \nu} + h_{\mu \nu} )} \left[-2\Lambda -\Box h - \frac{1}{6 m^2} \left(\Box h\right)^2\right]\,,
\end{equation}
and using the condition (\ref{condition}) as a sort of Lagrange multiplier (see also \cite{sergei}),  it becomes
\begin{equation}
\mathcal{L} = \sqrt{- \mathrm{det}  (\eta_{\mu \nu} + h_{\mu \nu} )} \left[ m^2 h-2 \Lambda - \frac{m^2}{6} h^2\right] \,.
\label{lm2}
\end{equation}
Working out the square root up to the second order \footnote{We expand $\sqrt{- \mathrm{det}  (\eta_{\mu \nu} + h_{\mu \nu} )}$ at the second order in  $h_{\mu \nu}$, in agreement with the order of expansion of $f(R)$ in $R$.} in $h_{\mu \nu}$, we find
\begin{equation}
\sqrt{- \mathrm{det}  (\eta_{\mu \nu} + h_{\mu \nu} )} \simeq 1+ \frac{1}{2} h+ \frac{1}{8} h^2- \frac{1}{4} h_{\mu \nu}h^{\mu \nu}\,,
\end{equation}
and the Lagrangian becomes
\begin{eqnarray}
\mathcal{L} &=& \left(1+ \frac{1}{2} h+ \frac{1}{8} h^2- \frac{1}{4} h_{\mu \nu}h^{\mu \nu}\right) \left(-2
\Lambda + m^2  h-\frac{m^2}{6} h^2\right)\nonumber\\&&
= - 2 \Lambda+ (m^2-\Lambda) h + \left(\frac{m^2}{3}-\frac{\Lambda}{4}\right) h^2 +
\frac{1}{2} \; \Lambda \;  h_{\mu \nu}h^{\mu \nu} + \frac{m^2}{24} h^3 - \frac{m^2}{48} h^4  \nonumber\\&&
 - \frac{m^2}{4} h \;h_{\mu \nu}h^{\mu \nu} +\frac{m^2}{24} h^2 \; h_{\mu \nu}h^{\mu \nu} \,.
\end{eqnarray}
By truncating up to the second order in $h$,  we get
\begin{equation}
\mathcal{L} =  -2 \Lambda+ (m^2-\Lambda) h + \left(\frac{m^2}{3}-\frac{\Lambda}{4}\right) h^2 +\frac{1}{2}
\; \Lambda \;  h_{\mu \nu}h^{\mu \nu}
\label{lagrangiana:massiva}
\end{equation}
This is a Lagrangian that  describe  a spin-0  particle
and a spin-2  particle.
The term proportional to $h$ does not affect the calculation of  perturbative observables, since it is linear in the creation and destruction operators. It vanishes when it is inserted between vacuum states.

\section{The effective  field  Lagrangian}

Eqs. \eqref{EFT33} and \eqref{lagrangiana:massiva} can be considered as effective Lagrangians written  in different variables.
In  \cite{Donoghue:2012zc},  the EFT is used to select the low energy modes, that are those of the GR, and contributions from quantum physics are analyzed. As in Eq. \eqref{EFT33}, the gravitational action is chosen proportional to powers of  the curvature $R$, but the arbitrary motivation of this choice is the physical smallness of $R$.
On the other hand, the expansion in $R$ comes out  naturally  from the ETG where the coefficients are fixed  from the effective Lagrangian.

Let us compare the effective Lagrangian  from linearized $f(R)$ gravity Eq.  \eqref{lagrangiana:massiva} with a Lagrangian derived on purely
 phenomenologically  ground.
The free part of the  Lagrangian for a massless spin-2 field can be written as
\begin{eqnarray}
{\cal L}_0= \frac{1}{2} \partial^\lambda (h_{\lambda \mu}+h_{ \mu \lambda}) \partial^\mu h -
 \frac{1}{4} \partial^\lambda \left(h_{\lambda \mu}+h_{ \mu \lambda}) \partial_\nu (h^{ \mu \nu}+ h^{\nu\mu} \right) + \nonumber \\
+ \frac{1}{8} \partial_\lambda (h_{ \mu \nu}+h_{ \nu \mu}) \partial^\lambda (h^{ \mu \nu}+ h^{\nu\mu} )
- \frac{1}{2} \partial_\lambda h  \partial^\lambda h\,.
\label{pheno}
\end{eqnarray}
This form is derived on the basis of Lorentz  invariance  and  gauge transformations as
\beq
h_{\mu\nu} \to h_{\mu\nu} +\partial_\mu \xi_\nu + \partial_\nu \eta_\mu
\eeq
where
$\xi_\nu$ and $\eta_\mu$ are eight arbitrary functions. Generic  mass terms can be added
\beq  {\cal L}_m=-a_1 h^2 -a_2 h_{\mu\nu}  h^{\mu\nu} -a_3 h_{\mu\nu}  h^{\nu\mu}
\label{genrel33} \eeq
with $a_1$, $a_2$ and $a_3$  being arbitrary coefficients.
In our case, $h_{\nu\mu}$ is symmetric, therefore the second and the third term coincide, which is equivalent to set, for instance, $a_3=0$.
The   Lagrangian ${\cal L}_0+{\cal L}_m$ describes
 an effective  theory with two  particles of  0-spin and 2-spin,
 as the Lagrangian in Eq. \eqref{lagrangiana:massiva}.
%
It has been demonstrated that, when $a_2 \neq a_3$,  the condition of null divergence of $h$ is not generally  respected by the scalar field, resulting in negative energy, or indefinite metric,  which are not physically acceptable  \cite{vanDam:1970vg}.
In order to recover  null divergence,  the coefficients and  the masses of Eq. \eqref{genrel33}  need to respect fixed relations among them.

It is not obvious to build sensible descriptions of the  gravitational interaction with this characteristic.
A standard way  is to use
 the  Hilbert-Einstein action for the massless gravitational field together with mass terms respecting the symmetries leading to the correct Ward identities.
In \cite{vanDam:1970vg}, it is observed that
such mass terms do not respect the relations necessary to a physically acceptable  Lagrangian  ${\cal L}_0+{\cal L}_m$, and we are forced to conclude
 that this  description of massive gravity is not satisfactory.

In our case,
the effective Lagrangian \eqref{lagrangiana:massiva}
 evades the condition $a_2 \neq a_3$ assumed in \cite{vanDam:1970vg}.
%
In fact, the  Lagrangian contains, at leading order,  only terms proportional to powers of $h$, which correspond to  $a_2 = a_3=0$.
Additional contributions
%
 are suppressed by $\Lambda$, in the limit  $\Lambda \ll m^2$, which is the same limit where
  dynamics of the scalar is  described by the physical Klein Gordon equation
\eqref{condition}.  In other words, we can say that starting from an analytical $f(R)$ gravity model, it is quite natural to recover an EFT where massive modes  emerge at scalar and tensor levels.

\section{Conclusions}
The issue of the consistency of a field theory for massive gravitons can  be settled by extending the Einstein gravity through generic functions of curvature invariants.  The minimal extension is $f(R)$ gravity where the standard Hilbert-Einstein action, linear in the Ricci scalar $R$ is substituted by a generic function. From a dynamical viewpoint, this means that further degrees of freedom of  gravitational field have to be taken into account and the possibility of massive gravitons naturally comes out, at least,  as massive scalar modes.  

In this paper, we have confronted the principal features of the Lagrangian resulting from the linearization of $f(R)$ gravity with the ones of the  effective phenomenological Lagrangian previously discussed in \cite{vanDam:1970vg}.  The main result is that it is possible to obtain massive terms  which indeed emerge  naturally if one breaks spontaneously the diffeomorphism invariance of GR, and, in this case, for a certain interval of parameters, it is possible to evade  ghosts and discontinuities.

Furthermore, it is possible to  identify  a natural mass scale $m$ directly related to the expansion parameters of the theory.  This fact could avoid to fix {\it by hand} the graviton mass since it comes directly from the structure of the theory. Upper limits (or mass ranges) could directly come by experimental constraints (see e.g. \cite{DeLaurentis:2011tp}).   Finally, in the limit  $m \gg \Lambda$
(or the less  restrictive one   $m^2 \gg \Lambda$),
 the theory results naturally  regularized and the massive scalar satisfies a physically acceptable Klein Gordon equation.
In a forthcoming paper, we will extend this approach to more general theories involving also the other curvature invariants.
Quantitative constraints to the massive modes resulting from the present analysis will be included.

\section*{Acknowledgments}

SC and MDL   acknowledge partial support of INFN Sez. di Napoli (iniziative specifiche QGSKY and  TEONGRAV).
 MDL is supported by MIUR (PRIN 2009).
GR acknowledges partial support by MIUR under project 2010YJ2NYW. GR and MP   acknowledge partial support of INFN Sez. di Napoli (iniziativa specifica RM21).


\begin{thebibliography}{0}

\bibitem{Fierz:1939ix}
  M.~Fierz and W.~Pauli,
  On Relativistic Wave Equations For Particles Of Arbitrary Spin In An
  Electromagnetic Field,
 {\it  Proc.\ Roy.\ Soc.\ Lond.\ A} {\bf 173}, (1939), 211-232.

\bibitem{vanDam:1970vg}
H.~van Dam and M.~J.~G.~Veltman, Massive And Massless Yang-Mills
And Gravitational Fields, {\it Nucl.\ Phys.\ B} {\bf 22}, (1970), 397-411.

\bibitem{Zakharov}
V.~I.~Zakharov, Linearized Gravitation Theory and the Graviton
Mass, {\it JETP Lett.} {\bf 12},  (1970), 312-314.

\bibitem{Vainshtein1972}
  A.~I.~Vainshtein, To The Problem Of Nonvanishing Gravitation Mass,
  {\it Phys.\ Lett.\ B} {\bf 39}, (1972), 393-394.

\bibitem{Boulware:1973my}
D.~G.~Boulware and S.~Deser, Can Gravitation Have A Finite
Range?, {\it Phys.\ Rev.\ D} {\bf 6}, (1972), 3368-3382.

\bibitem{Arkani-Hamed:2002sp}
N.~Arkani-Hamed, H.~Georgi and M.~D.~Schwartz, Effective field
theory for massive gravitons and gravity in theory space, {\it Annals
Phys.\ } {\bf 305}, (2003), 96-118. 

\bibitem{Rubakov:2008nh}
  V.~A.~Rubakov and P.~G.~Tinyakov, Infrared-modified gravities and massive gravitons,
  {\it Phys.\ Usp.} {\bf 51},(2008) 759-792.
  
 \bibitem{hinterbichler}
 K. Hinterbichler, Theoretical Aspects of Massive Gravity, {\it Rev. Mod. Phys.}  {\bf 84},  (2012),  671-710.
 
\bibitem{wonlee}
H. W. Lee, K.Y. Kim, and Y. S. Myung, Massive gravitons dark matter scenario revisited, 
{\it Mod. Phys. Lett. A} { 27}, (2012), 1250146-1250159.

\bibitem{casadio}
L. Bellagamba, R. Casadio, R. Di Sipio, V. Viventi, Black Hole Remnants at the LHC,  
 {\it Eur.Phys.J.} {\bf C 72}, (2012), 1957-1970 .

   \bibitem{Smolin03} L. Smolin, How far are we from the quantum theory of gravity? (2003), arXiv:hep-th/0303185

\bibitem{Kiefer06}C.
Kiefer, Quantum gravity: General introduction and recent developments,  {\it Annalen der Physik}, {\bf 15}, (2006),129-148.

\bibitem{Will06}C.
Will, The Confrontation between general relativity and experiment, {\it Living Reviews in Relativity}, {\bf 9}, (2006), 3-89.



\bibitem{Visser98}M.
Visser, Mass for the graviton, {\it General Relativity and Gravitation}, 30, (1998), 1717-1728.

\bibitem{DDGV02}
C. Deffayet, G. Dvali, G. Gabadadze, A.Vainshtein,  Nonperturbative continuity in graviton mass versus perturbative discontinuity, {\it Phys.Rev. D}
  {\bf 65}, (2002), 044026-044045.

\bibitem{DGP00}
{Dvali}, G., {Gabadadze}, G., \& {Porrati}, M. , 4-D gravity on a brane in 5-D Minkowski space,  Physics Letters B, 485, (2000), 208-214.

\bibitem{PhysRepnostro} S. Capozziello, M. De Laurentis, Extended Theories of Gravity, {\it Phys. Rept.} {\bf 509}, (2011) 167-321.
\bibitem{OdintsovPR}S. Nojiri, S.D. Odintsov, Unified cosmic history in modified gravity: from F(R) theory to Lorentz non-invariant models, {\it Phys. Rept.} {\bf 505}, (2011) 59-144.
\bibitem{Mauro}S. Capozziello, M. Francaviglia, Extended Theories of Gravity and their Cosmological and Astrophysical Applications, {\it Gen. Rel. Grav.} {\bf 40}, (2008) 357-420. 
\bibitem{DeFelice} A. De Felice, S. Tsujikawa, f(R) theories, {\it Living Rev.Rel.} {\bf 13}, (2010), 3-139.

\bibitem{basini}S. Capozziello, G. Basini, M. De Laurentis, Deriving the mass of particles from Extended Theories of Gravity in LHC era, 
 {\it Eur. Phys. Jou. C} {\bf  71}, (2011) 1679-1709.
  
\bibitem{riemann}
G. F. B. Riemann,G.F.B., {\it Uber dieHypothesen, welche der Geometrie zu Grunde liegen }, Abhand. K. Ges. Wiss. Gottingen, {\bf 13}, 133 (1868);
English translation by Clifford, W. K. {\it Nature} {\bf 8}, 14, (1873); reprinted and edited by Weyl, H., Springer, Berlin, 1920. Included in its {\it Gesammelte Mathematische Werke, wissenschaftlicher Nachlass und Nachtrage}, eds. Weber, H., Dedekind, R., Teubner, B. G., Leipzig, (1892); 2d ed. Dover Publ., New York.

\bibitem{greci}C. Bogdanos, S. Capozziello, M. De Laurentis, S. Nesseris, Massive, massless and ghost modes of gravitational waves from higher-order gravity, {\it Astrop. Phys.}{\bf 34},  (2010) 236-244.

\bibitem{Hare73}
M. G. Hare, Mass of the graviton, {\it Canadian Journal of Physics}, 51, (1973) 431-433.

\bibitem{GN74}
A.S. Goldhaber, M.M. Nieto, Mass of the graviton, {\it Phys.Rev. D} {\bf 9} (1974) 1119-1121.

\bibitem{FS02}L.S. Finn, P.J. Sutton,  Bounding the mass of the graviton using binary pulsar observations, {\it Phys.Rev. D} {\bf 65} (2002) 044022-044037.

\bibitem{ivan} M. De Laurentis, I. De Martino, Testing f(R) theories using the first time derivative of the orbital period of the binary pulsars, {\it Mon. Not. Roy. Astr. Soc.} {\bf 431} (1), (2013), 741-748.

\bibitem{Donoghue:1995cz}
  J.~F.~Donoghue, Introduction to the effective field theory description of gravity, (1995)
  gr-qc/9512024.
  
\bibitem{Burgess:2003jk}
  C.~P.~Burgess, Quantum gravity in everyday life: General relativity as an effective field theory, {\it Living Rev.\ Rel.} {\bf 7} (2004) 5-62. 
  
  
\bibitem{Donoghue:2012zc}
  J.~F.~Donoghue,The effective field theory treatment of quantum gravity,
  {\it AIP Conf.\ Proc.} {\bf 1483} (2012) 73-94  
  
\bibitem{starobinsky}
 A.A. Starobinsky,  A New Type of Isotropic Cosmological Models Without Singularity,  {\it Phys. Lett.} {\bf B 91}, (1980), 99-102.

\bibitem{Maggiore}
M.~Maggiore, {\it Gravitational Waves: Theory and Experiments}
(Oxford Univ. Press, Oxford, 2007).

\bibitem{DeLaurentis:2011re}
M. De Laurentis, Newtonian and relativistic theory of orbits and emission of gravitational waves, {\it The Open Astronomy Journal} {\bf 4}, (2011), 1874-1918 .


\bibitem{DeLaurentis:2011tp}
  M.~De Laurentis and S.~Capozziello, Quadrupolar gravitational radiation as a test-bed for f(R)-gravity, {\it Astropart.\ Phys.}  {\bf 35}, (2011), 257-265.

\bibitem{sergei}
S. Capozziello, J. Matsumoto, S. Nojiri, S. D. Odintsov, Dark energy from modified gravity with Lagrange multipliers, {\it Phys. Lett. B} {\bf 693}, (2010), 198-208.


\end{thebibliography}
\end{document}